\documentclass[12pt]{article}

\usepackage[english]{babel}
\usepackage[usenames]{color}
\usepackage[cp1250]{inputenc}
\usepackage{amsfonts}
\usepackage{amsthm}
\usepackage{graphicx}
\usepackage{epsfig}
\usepackage{mathrsfs}
\usepackage{amsmath}

\theoremstyle{plain}

\begin{document}
\newcommand{\bea}{\begin{eqnarray}}
\newcommand{\eea}{\end{eqnarray}}
\newcommand{\be}{\begin{equation}}
\newcommand{\ee}{\end{equation}}
\newcommand{\beas}{\begin{eqnarray*}}
\newcommand{\eeas}{\end{eqnarray*}}
\newcommand{\bs}{\backslash}
\newcommand{\bc}{\begin{center}}
\newcommand{\ec}{\end{center}}
\def\SC {\mathscr{C}}

\title{Asymmetric numeral systems:\\ entropy coding combining speed of Huffman coding \\ with compression rate of arithmetic coding}
\author{Jarek Duda}
\date{\it \footnotesize Center for Science of Information, Purdue University, W. Lafayette, IN 47907, U.S.A.\\
\textit{email:} dudaj@purdue.edu}
\maketitle

\begin{abstract}
The modern data compression is mainly based on two approaches to entropy coding: Huffman (HC) and arithmetic/range coding (AC). The former is much faster, but approximates probabilities with powers of 2, usually leading to relatively low compression rates. The latter uses nearly exact probabilities - easily approaching theoretical compression rate limit (Shannon entropy), but at cost of much larger computational cost.

Asymmetric numeral systems (ANS) is a new approach to accurate entropy coding, which allows to end this tradeoff between speed and rate: the recent implementation \cite{fse} provides about 50\% faster decoding than HC for 256 size alphabet, with compression rate similar to provided by AC. This advantage is due to being simpler than AC: using single natural number as the state, instead of two to represent a range. Beside simplifying renormalization, it allows to put the entire behavior for given probability distribution into a relatively small table: defining entropy coding automaton. The memory cost of such table for 256 size alphabet is a few kilobytes. There is a large freedom while choosing a specific table - using pseudorandom number generator initialized with cryptographic key for this purpose allows to simultaneously encrypt the data.

This article also introduces and discusses many other variants of this new entropy coding approach, which can provide direct alternatives for standard AC, for large alphabet range coding, or for approximated quasi arithmetic coding.
\end{abstract}
\section{Introduction}
In standard numeral systems different digits as treated as containing the same amount of information: 1 bit in the binary system, or generally $\lg(b)$ bits ($\lg\equiv \log_2$) in base $b$ numeral system. However, event of probability $p$ contains $\lg(1/p)$ bits of information. So while standard numeral systems are optimal for uniform digit probability distributions, to optimally encode general distributions, what is the heart of data compression, we should try to somehow asymmetrize this concept. We can obtain arithmetic/range coding(AC) (\cite{ari}, \cite{ran}) or recent asymmetric numeral system(ANS) (\cite{me}, \cite{ans}) this way, depending on position we add succeeding digits.

Specifically, in standard binary numeral system, having information stored in a natural number $x$ (having $m$ digits), we can add information from a digit $s\in\{0,1\}$ in basically two different ways: in the most significant position ($x \rightarrow x+ 2^m s$) or in the least significant position ($x \rightarrow 2x+s$). The former means that the new digit chooses between large ranges - we can asymmetrize it by changing proportions of these ranges, getting arithmetic/range coding. However, while in standard numeral systems we would need to remember position of this most significant digit ($m$), AC requires to specify the range in given moment - its current state are two numbers representing the range.

\begin{figure}[b!]
    \centering
        \includegraphics{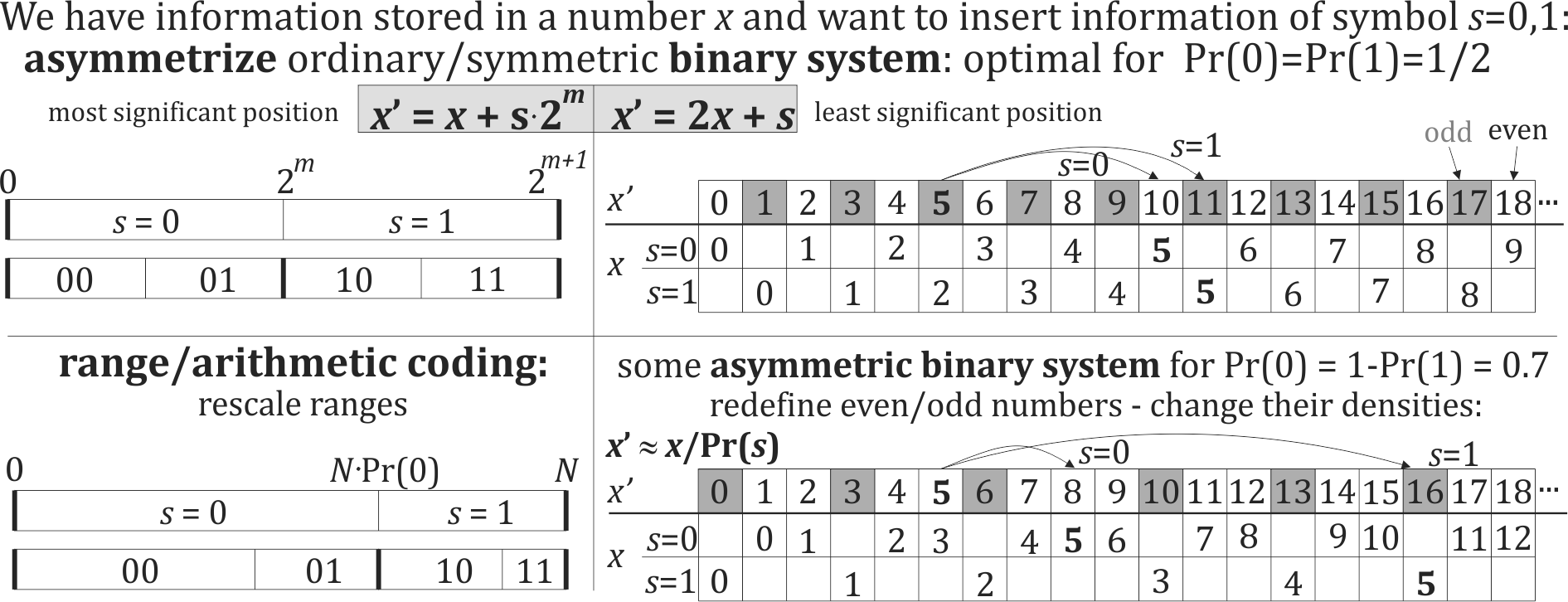}\
        \caption{Two ways to asymmetrize binary numeral system. Having some information stored in a natural number $x$, to attach information from 0/1 symbol $s$, we can add it in the most significant position ($x'=x+s 2^m$), where $s$ chooses between ranges, or in the least significant ($x'=2x+s$) position, where $s$ chooses between even and odd numbers. The former asymmetrizes to AC by changing range proportions. The latter asymmetrizes to ABS by redefining even/odd numbers, such that they are still uniformly distributed, but with different density. Now $x'$ is $x$-th element of the $s$-th subset. }
        \label{intr}
\end{figure}

In contrast, while adding information in the least significant position, the current state is just a single natural number. This advantage remains for the asymmetrization: ANS. While in standard binary system $x$ becomes $x$-th appearance of corresponding even ($s=0$) or odd ($s=1$) number, this time we would like to redefine this splitting of $\mathbb{N}$ into even and odd numbers. Such that they are still "uniformly distributed", but with different densities - corresponding to symbol probability distribution we would like to encode, like in Fig. \ref{intr}.

Let us look at both approaches from information theory point of view. For AC, knowing that we are currently in given half of the range is worth 1 bit of information, so the current content is lg(size of range/size of subrange) bits of information. While encoding symbol of probability $p$, the size of subrange is multiplied by $p$, increasing informational content by $\lg(1/p)$ bits as expected. From ANS point of view, seeing $x$ as containing $\lg(x)$ bits of information, informational content should increase to $\lg(x)+\lg(1/p)=\lg(x/p)$ bits while adding symbol of probability $p$. So we need to ensure $x\rightarrow \approx x/p$ transition relation. Intuitively, rule: $x$ becomes $x$-th appearance of $s$-th subset fulfills this relation if this subset is uniformly distributed with $\Pr(s)$ density.\\

\begin{figure}[t!]
    \centering
        \includegraphics{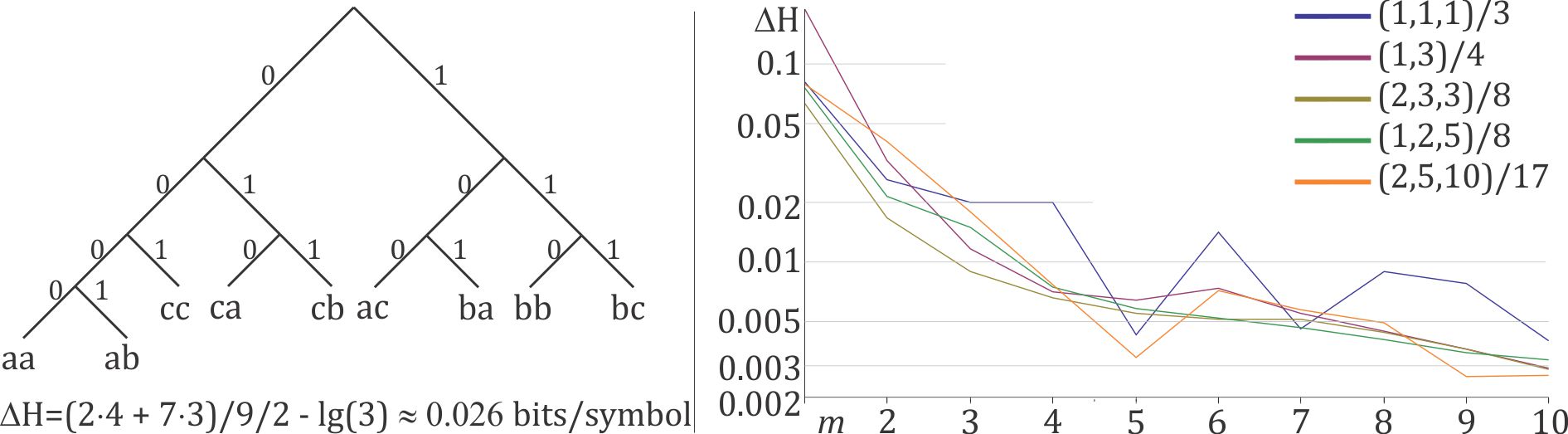}\
        \caption{Left: construction of Huffman coding while grouping two symbols from $(1,1,1)/3$ probability distribution: we have 9 equiprobable possibilities. This grouping reduces distance from the Shannon entropy: $\Delta H$. Standard base 3 numeral system would give $\Delta H=0$ here. Right: $\Delta H$ for grouping $m$ symbols. For example operating on $3^{10}=59049$ possibilities allows to get loss $\Delta H\approx 0.003$ bits/symbol. In comparison, $\Delta H$ for ANS using just 16 or 17 states for the last four cases is correspondingly: $\approx$ 0.00065, 0.00122, 0.00147, 0.00121 bits/symbol.}
        \label{huffman}
\end{figure}

Huffman coding \cite{huf} can be seen as a memoryless coder: type "symbol$\rightarrow$ bit sequence" set of rules. It transforms every symbol into a natural number of bits, approximating their probabilities with powers of 1/2. Theoretically it allows to approach the capacity (Shannon entropy) as close as we want by grouping multiple symbols. However, we can see in Fig. \ref{huffman} that this convergence is relatively slow: getting $\Delta H\approx$0.001 bits/symbol would be completely impractical here. Precise analysis can be found in \cite{huff}.

Let us generally look at the cost of approximating probabilities. If the we are using a coder which encodes perfectly $(q_s)$ symbol distribution to encode $(p_s)$ symbol sequence, we would use on average $\sum_s
p_s \lg(1/q_s)$ bits per symbol, while there is only Shannon entropy needed: $\sum_s p_s \lg(1/p_s)$. The difference between them is called Kullback - Leiber distance:
\be \Delta H=\sum_s p_s \lg\left(\frac{p_s}{q_s}\right)\approx \sum_s\frac{-p_s}{\ln(2)}\left(\left(1-\frac{q_s}{p_s}\right)-
\frac{1}{2}\left(1-\frac{q_s}{p_s}\right)^2\right)\approx 0.72\sum_s \frac{(\epsilon_s)^2}{p_s} \label{KL}\ee
where $\epsilon_s=q_s-p_s$ will be referred as \emph{inaccuracy}.

So for better than Huffman convergence, we need more accurate coders - we need to handle fractional numbers of bits. It can be done by adding a memory/buffer to the coder, containing noninteger number of bits. The coder becomes a finite state automaton, with
$$\textrm{type \qquad\quad(symbol, state)}\rightarrow \textrm{(bit sequence, new state)}\qquad\quad \textrm{coding rules.}$$ 

\begin{figure}[t!]
    \centering
        \includegraphics{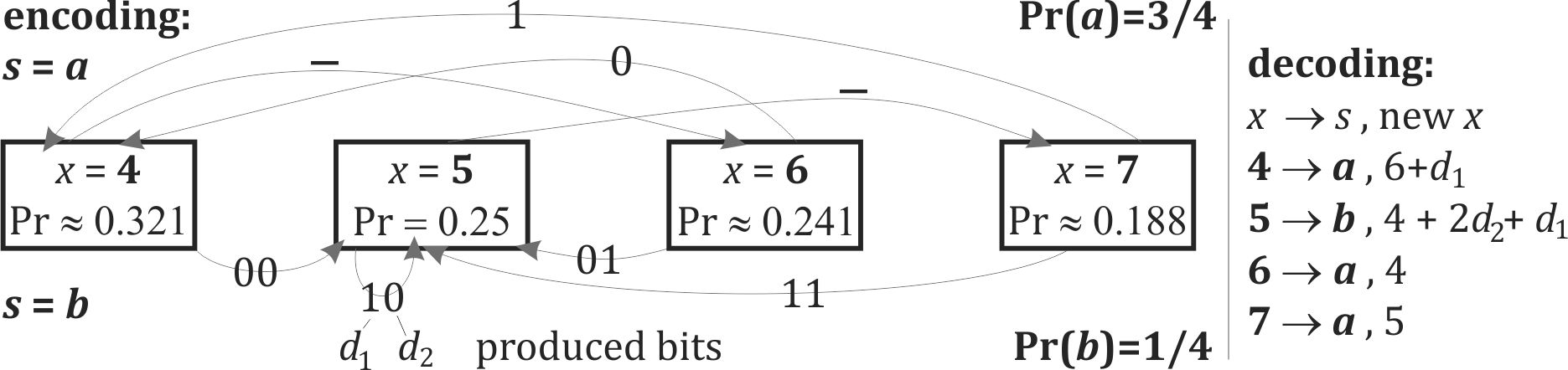}\
        \caption{Some 4 state encoding and decoding automaton for ANS with $\Pr(a)=3/4,\ \Pr(b)=1/4$ probability distribution. Upper edges of encoding picture are transitions for symbol "$a$", lower for symbol "$b$". Some edges contain digits produced while corresponding transition. Intuitively, $x$ is a buffer containing $\lg(x)\in[2,3)$ bits of information - symbol "$b$" always produces 2 bits, while "$a$" accumulates in the buffer. Decoding is unique because each state corresponds to a fixed symbol and number of digits to process: 1 for state 4, 2 for state 5, none for 6, 7. There is written stationary probability distribution for i.i.d. source, which allows to find average number of used bits per symbol: $\approx 2\cdot 1/4 + 1\cdot 3/4\cdot(0.241+0.188)\approx 0.82$, what is larger than Shannon entropy by $\Delta H\approx 0.01$ bits/symbol. Increasing the number of states to 8 allows to reduce it to $\Delta H\approx 0.0018$ bits/symbol. }
        \label{auto}
\end{figure}

As it was discussed, the current state contains lg(size of range/size of subrange) bits in the case of AC, or $\lg(x)$ bits in the case of ANS. Allowing such state to contain large number of bits would require to operate with large precision in AC, or large arithmetic in ANS - is impractical. To prevent that, in AC we regularly perform renormalization: if the subrange is in one half of the range, we can send single bit to the stream (pointing this half), and rescale this half up to the whole range. Analogously we need to gather accumulated bits in ANS: transfer to the stream some least significant bits of the state, such that we will return to a fixed range ($I=\{l,..,2l-1\}$) after encoding given symbol. In contrast to AC, the number of such bits can be easily determined here. Example of such 4 state ANS based automaton for $I=\{4,5,6,7\}$ can be see in Fig. \ref{auto}: while symbol $b$ of 1/4 probability always produces 2 bits of information, symbol "a" of 3/4 probability usually accumulates information by increasing the state, finally producing complete bit of information. While decoding, every state know the number of bits to use.\\

This article introduces and discusses many variants of applying the ANS approach or its binary case: asymmetric binary system (ABS). Most of them have nearly a direct alternative in arithmetic coding family:\\

\begin{tabular}{c|c|c|}
   & AC & ANS \\ \hline
 very accurate, formula for binary alphabet & AC & uABS, rABS \\
 fast, tabled for binary alphabet & M coder & tABS \\
 covering all probabilities with small automata & quasi AC & qABS \\
 very accurate, formula for large alphabet & Range Coding & rANS \\
 tabled for large alphabet  & - & tANS \\
\end{tabular} \\

There will be now briefly described these variants. The initial advantage of ANS approach is much simpler renormalization. In AC it is usually required to use slow branches to extract single bits and additionally there is an issue when range contains the middle point. These computationally costly problems disappear in ANS: we can quickly deduce the number of bits to use in given step from $x$, or store them in the table - we just directly transfer the whole block of bits once per step, making it faster.

uABS stands for direct arithmetic formula for \textbf{u}niformly distributing symbols. rABS stands for distributing symbols in \textbf{r}anges - leading to still direct formula, a bit less accurate but simpler to calculate. Its main advantage is allowing for large alphabet version: rANS, which can be seen as direct alternative for Range Coding, but with a small advantage: instead of 2 multiplications per step, it requires only a single one. We can also put e.g. uABS behavior into tables, getting tABS. It can be applied similarly like one of many AC approximations, for example M coder used in CABAC \cite{CABAC} in modern video compression. Matt Mahoney implementations of tABS (fpaqb) and uABS (fpaqc) are available in \cite{mah}.

While those very accurate versions can easily work loss many orders of magnitude below $\Delta H\approx$0.001 bits/symbol, qABS resembles \textbf{q}uasi arithmetic coding \cite{quasi} approximation: we would like to construct a family of small entropy coding automata, such we can cover the whole space of (binary) probability distributions by switching between them. For qABS it is 5 automata with 5 states for $\Delta H\approx$0.01 bits/symbol loss, or about 20 automata with 16 states for $\Delta H\approx$0.001 bits/symbol.

While above possibilities might not bring a really essential advantage, due to much smaller state space, \textbf{t}abled ANS puts the whole behavior for large alphabet into a relatively small coding table, what would be rather too demanding for AC approach. There is available interactive demonstration of this tANS: \cite{dem}. The number of states of such automaton should be about 2-4 times larger than the size of alphabet for $\Delta H\approx$0.01 bits/symbol loss, or about 8-16 times for $\Delta H\approx$0.001 bits/symbol. For 256 size alphabet it means 1-16kB memory cost per such coding table. Generally $\Delta H$ drops with the square of the number of states: doubling the number of states means about 4 times smaller $\Delta H$. As in the title, in this way we obtain extremely fast entropy coding: recent Yann Collet implementation \cite{fse} has 50\% faster decoding than fast implementation of Huffman coding, with nearly optimal compression rate - combining advantages of Huffman and arithmetic coding.

There is a large freedom while choosing the coding table for given parameters: every symbol distribution defines an essentially different coding. We will also discuss three reasons of chaotic behavior of this entropy coder: asymmetry, ergodicity and diffusion, making extremely difficult any trial of tracing the state without complete knowledge. These two reasons suggest to use it also for cryptographic applications: use pseudorandom number generator initialized with a cryptographic key to choose the exact simbol distribution/coding. This way we simultaneously get a decent encryption of encoded data.
\section{Basic concepts and versions} \label{abs}
We will now introduce basic concepts for encoding information in natural numbers and find analytic formulas for uABS, rABS and rANS cases.
\subsection{Basic concepts}
There is given an alphabet $\mathcal{A}=\{0,..,n-1\}$ and assumed probability distribution $\{p_s\}_{s\in \mathcal{A}}$, $\sum_s p_s=1$. The \emph{state} $x\in \mathbb{N}$ in this section will contain the entire already processed symbol sequence.

We need to find \emph{encoding} ($C$) and \emph{decoding} ($D$) \emph{functions}. The former takes a state $x\in\mathbb{N}$ and symbol $s\in \mathcal{A}$, and transforms them into $x'\in \mathbb{N}$ storing information from both of them. Seeing $x$ as a possibility of choosing a number from $\{0,1,..,x-1\}$ interval, it contains $\lg(x)$ bits of information. Symbol $s$ of probability $p_s$ contains $\lg(1/p_s)$ bits of information, so $x'$ should contain approximately $\lg(x)+\lg(1/p_s)=\lg(x/p_s)$ bits of information: $x'$ should be approximately $x/p_s$, allowing to choose a value from a larger interval $\{0,1,..,x'-1\}$. Finally we will have functions defining single steps:
$$ C(s,x)=x',\quad D(x')=(s,x)\quad:\quad D(C(s,x))=(s,x),\quad C(D(x'))=x',\quad x'\approx x/p_s$$

For standard binary system we have $C(s,x)=2x+s$, $D(x')=(\textrm{mod}(x,2), \lfloor x/2\rfloor)$ and we can see $s$ as choosing between even and odd numbers. If we imagine that $x$ chooses between even (or odd) numbers in some interval, $x'=2x+s$ chooses between all numbers in this interval.

For the general case, we will have to redefine the subsets corresponding to different $s$: like even/odd numbers they should still uniformly cover $\mathbb{N}$, but this time with different densities: $\{p_s\}_{s\in \mathcal{A}}$. We can define this split of $\mathbb{N}$ by a \emph{symbol distribution} $\overline{s}: \mathbb{N}\to \mathcal{A}$
$$\{0,1,..,x'-1\}=\bigcup_s \left\{x\in\{0,1,..,x'-1\}:\overline{s}(x)=s\right\}$$
While in standard binary system $x'$ is $x$-th appearance of even/odd number, this time it will be $x$-th appearance of $s$-th subset. So the decoding function will be
\be D(x)=(\overline{s}(x),x_{\overline{s}(x)})\qquad \textrm{where}\qquad x_s:=\left|\{y\in\{0,1,..,x-1\}:\overline{s}(y)=s\}\right| \ee
and $C(s,x_s)=x$ is its inversion. Obviously we have $x=\sum_s x_s$.

As $x/x_s$ is the number of bits we currently use to encode symbol $s$, to reduce inaccuracy and so $\Delta H$, we would like that $x_s\approx xp_s$ approximation is as close as possible - what intuitively means that symbols are nearly uniformly distributed with $\{p_s\}$ density. We will now find formulas for the binary case by just taking $x_1:=\lceil xp \rceil$ and in Section \ref{ans} we will focus on finding such nearly uniform distributions on a fixed interval for larger alphabets.\\

Finally we can imagine that $x$ is a stack of symbols, $C$ is push operation, $D$ is pop operation. The current encoding algorithm would be: start e.g. with $x=1$ and then use $C$ with succeeding symbols. It would lead to a large natural number, from which we can extract all the symbols in reversed order using $D$. To prevent inconvenient operations on large numbers, in the next section we will discuss stream version, in which cumulated complete bits will be extracted to make that $x$ remains in a fixed interval $I$.
\subsection{Uniform asymmetric binary systems (uABS)}\label{abss}
We will now find some explicit formulas for the binary case and nearly uniform distribution of symbols: $\mathcal{A}=\{0,1\}$.

Denote $p:=p_1$, $\tilde p:=1-p=p_0$. To obtain $x_s\approx x\cdot p_s$ we can for example choose
\be x_1:=\lceil xp \rceil\quad\quad\quad\quad\quad\quad\quad\left(\textrm{or alternatively}\ x_1:=\lfloor xp \rfloor\right) \ee
\be x_0=x-x_1=x-\lceil xp \rceil \quad\quad\quad\quad\quad\left(\textrm{or}\ \ x_0=x-\lfloor xp \rfloor \right)\ee
Now $\overline{s}(x)=1$ if there is a jump of $\lceil xq \rceil$ in the succeeding position:
\be s:=\lceil (x+1)p \rceil-\lceil xp \rceil \qquad\qquad\qquad\left(\textrm{or}\ s:=\lfloor (x+1)p \rfloor-\lfloor xq \rfloor\right) \label{abs1} \ee
This way we have found some \textbf{decoding} function: $D(x)=(s,x_s)$.

We will now find the corresponding encoding function: for given $s$ and $x_s$ we want to find $x$. Denote $r:=\lceil xq \rceil-xq\in[0,1)$\\
\be \overline{s}(x)=s=\lceil (x+1)p \rceil-\lceil xp \rceil=\lceil (x+1)q -\lceil xq \rceil\rceil=\lceil (x+1)p-r-xq\rceil=\lceil p-r \rceil\ee
$$s=1\Leftrightarrow r<p$$
\begin{itemize}
\item $s=1$: $\quad x_1=\lceil xp \rceil=xp+r$\\
$x=\frac{x_1-r}{p}=\Big\lfloor\frac{x_1}{p}\Big\rfloor\quad$
as it is a natural number and $0\leq r<p$.
\item $s=0$: $p\leq r<1$ so $\tilde{p}\geq 1-r>0$\\
$x_0=x-\lceil xp \rceil=x-xp-r=x\tilde{p}-r$
$$x=\frac{x_0+r}{\tilde{p}}=\frac{x_0+1}{\tilde{p}}-\frac{1-r}{\tilde{p}}=
\Big\lceil\frac{x_0+1}{\tilde{p}}\Big\rceil-1$$
\end{itemize}
Finally \textbf{encoding} is:
\be \label{coding} C(s,x)=\left\{\begin{array}{ll}
\Big\lceil\frac{x+1}{1-p}\Big\rceil-1 &\ \textrm{if}\ s=0\\
\ \Big\lfloor\frac{x}{p}\Big\rfloor &\ \textrm{if}\ s=1\end{array}
\right. \quad\quad\quad\left(\textrm{or}\ = \left\{
\begin{array}{ll}
\ \ \ \Big\lfloor\frac{x}{1-p}\Big\rfloor &\ \textrm{if}\ s=0\\
\Big\lceil\frac{x+1}{p}\Big\rceil-1 &\ \textrm{if}\ s=1\end{array}
\right.\right)\ee
For $p=1/2$ it is the standard binary numeral system with switched digits.\\

The starting values for this formula and $p=0.3$ are presented in bottom-right of Fig. \ref{intr}. Here is an example of encoding process by inserting succeeding symbols:
\be 1\xrightarrow{1}3\xrightarrow{0} 5\xrightarrow {0} 8\xrightarrow{1} 26\xrightarrow{0} 38\xrightarrow{1} 128 \xrightarrow{0} 184\xrightarrow{0}264 ...\label{ex1}\ee

We could directly encode the final $x$ using $\lfloor \lg(x) \rfloor +1$ bits. Let us look at the growth of $\lg(x)$ while encoding: symbol $s$ transforms state from $x_s$ to $x$:
$$-\lg\left(\frac{x_s}{x}\right)=-\lg\left(p_s+\epsilon_s(x)\right)=-\lg(p_{s}) -\frac{\epsilon_s(x)}{p_s\ln(2)}+O((\epsilon_s(x))^2)$$
where $$\epsilon_s(x)=x_s/x-p_s \qquad\qquad \textrm{describes inaccuracy}.$$
In the found coding we have $|x_s-xp_s|<1$ and so $|\epsilon_s(x)|<1/x$. While encoding symbol sequence of $\{p_s\}_{s\in\mathcal{A}}$ probability distribution, the sum of above expansion says that we need on average $H=-\sum_s p_s \lg(p_s)$ bits/symbol plus higher order terms. As $x$ grows exponentially while encoding, $|\epsilon_s(x)|<1/x$, so these corrections are $O(1)$ for the whole sequence.
\subsection{Range variants (rABS, rANS)}
We will now introduce alternative approach, in which we place symbol appearances in ranges. It is less accurate, but can be less expensive to perform. Most importantly, it allows for large alphabets, getting direct alternative for Range Coding - this time using single multiplication per symbol instead of two required in Range Coding. This approach can be seen as taking a standard numeral system and merging some of its succeeding digits.\\

We will start with binary example (rABS), and then define general formulas.\\
\textbf{Example:} Looking at (1,3)/4 probability distribution, we would like to take a standard base 4 numeral system: using digits 0, 1, 2, 3 and just merge 3 of its digits - let say 1, 2 and 3. So while the symbol distribution for standard ternary system would be $\overline{s}(x)=\textrm{mod}(x,4)$: cyclic (0123), we will asymmetrize it to (0111) cyclic distribution:
$$\overline{s}(x)=0 \quad\textrm{if}\quad\textrm{mod}(x,4)=0 \quad \textrm{else}\quad \overline{s}(x)=1$$
Behavior for $s=0$ is exactly like in ternary system. In contrast, for $s=1$ we will go to $\lfloor x/3\rfloor$-th of (0111) quadruples:
$$C(0,x)=4x\qquad\qquad C(1,x)=4 \lfloor x/3\rfloor + \textrm{mod}(x,3)+1$$
where the "+1" is to shift over the single "0" appearance. Decoding:
$$\textrm{if}(\textrm{mod}(x,4)=0)\quad \{s=0;\ x=\lfloor x/4\rfloor\} \quad\textrm{else}\quad \{s=1;\ x=3\lfloor x/4\rfloor + \textrm{mod}(x,4)-1\}$$
Analogously we can define coding/decoding functions for different fractions, what is a special case of large alphabet formulas we will find now: rANS. Assume that the probability distribution is a fraction of form: $(l_0,l_1,...,l_{n-1})/m$ where $l_s\in \mathbb{N},\ m=\sum_s l_s$.

Now let us analogously imagine that we start with base $m$ numeral system and merge corresponding numbers of succeeding digits together: the symbol distribution is cyclic (00..011..1..."n-1") with $l_s$ appearances of symbol $s$. Let us define $s(x)$ as symbol in $x\in [0,..,m-1]$ position in this cycle: 
$$\overline{s}(x)=s(\textrm{mod}(x,m))\qquad \textrm{where}\quad s(x) = \min \left\{s: x < \sum_{i=0}^s l_i\right\}$$
and $b_s:=\sum_{i=0}^{s-1} l_i$ is the starting position of $s$ symbol in this cycle. These $l_s$, $b_s$ and $s(x)$ should be tabled in coder. Putting $l(x)=l_{s(x)}$ and $b(x)=b_{s(x)}$ into tables would reduce the number of use per step to one.

Now encoding and decoding steps are:
$$C(s,x)=m\lfloor x/l_s \rfloor + b_s +\textrm{mod}(x,l_s)$$
$$D(x)= (s,\ l_s \lfloor x/m\rfloor + \textrm{mod}(x,m) -b_s)\qquad\textrm{where}\qquad s=s(\textrm{mod}(x,m))$$
If we choose $m$ as a power of 2, multiplying and dividing by it can be made by bit shifts, $\textrm{mod}(x,m)$ by bitand with mask - decoding requires only single multiplication per step. In contrast, in Range Coding there are needed two - this approach should be faster.\\

Let us find some boundary of its inaccuracy $\epsilon_s(x)=x_s/x-p_s$ :
$$|x_s/x-p_s|=\left|\frac{l_s \lfloor x/m\rfloor + \textrm{mod}(x,m) -b_s}{x}-\frac{l_s}{m}\right|=\left|\frac{(1-l_s/m)\textrm{mod}(x,m) -b_s}{x}\right|<m/x$$
as $\lfloor x/m\rfloor =(x-\textrm{mod}(x,m))/m$ and $b_s<m$.

This inaccuracy boundary is $m$ times larger than for uABS. However, as this approach is intended to directly apply above arithmetic formulas, we can use relatively large $x$, making this inaccuracy negligible.
\section{Stream version - encoding finite-state automaton}                   \label{str}
\begin{figure}[t!]
    \centering
        \includegraphics{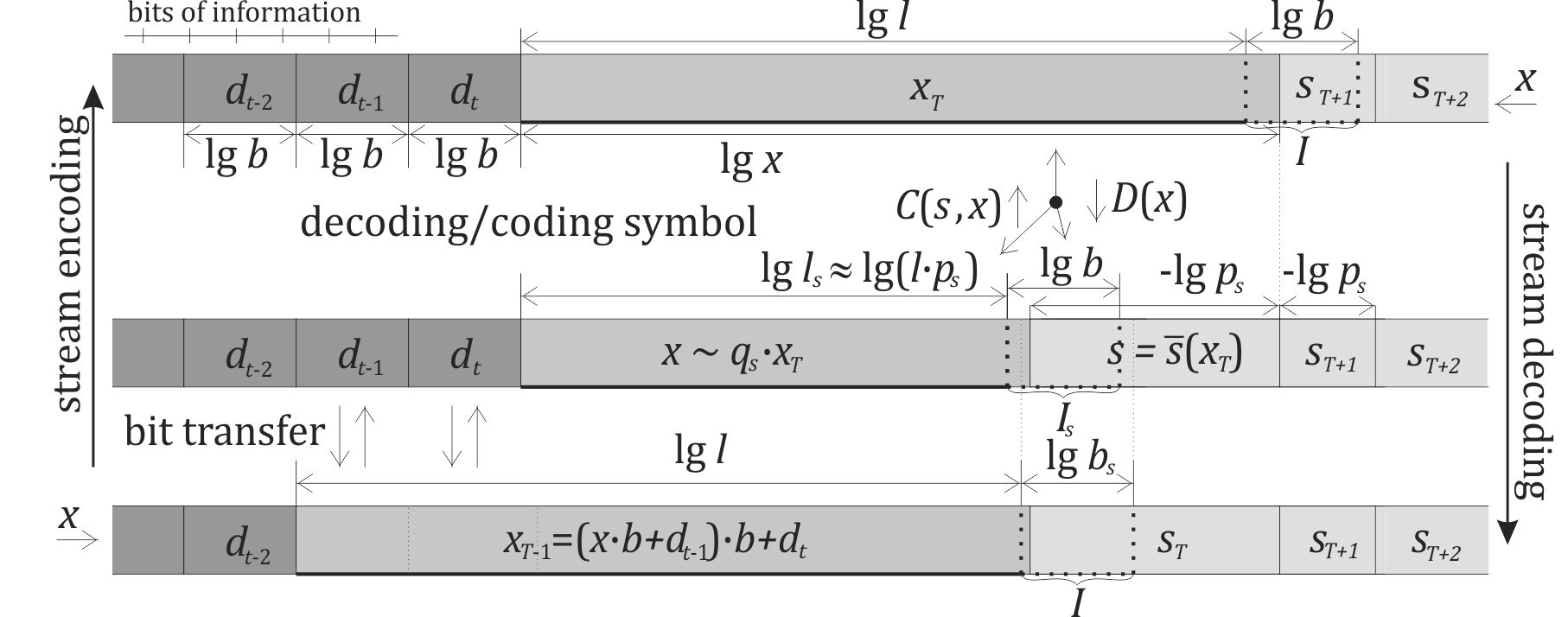}\
        \caption{Schematic picture of stream encoding/decoding: encoder travels right, transforming symbols containing $\lg(1/p_s)$ bits of information, into digits containing $\lg(d)$ bits each. When information accumulates in buffer $x$, digits are produced such that $x$ remains in $I={l,..,bl-1}$. Decoder analogously travels left.}
        \label{stream}
\end{figure}

Arithmetic coding requires renormalization to work with finite precision. Analogously to prevent ANS state from growing to infinity, we will enforce $x$ to remain in some chosen range $I=\{l,..,2l-1\}$. This way $x$ can be seen as a buffer containing between $l$ and $l+1$ bits of information. So while we operate on symbols containing non-integer number of bits, as they accumulate inside the buffer, we can extract complete bits of information to remain in its operating region. These extracted bits are transferred to the bitstream and used while decoding. Finally the situation will intuitively look like in Fig. \ref{stream}.
\subsection{Algorithm}  \label{alg}
To make these considerations more general, we will extract digits in base $2\leq b\in \mathbb{N}$ numeral system. Usually we will use bits: $b=2$, but sometimes using a larger $b$ could be more convenient, for example $b=2^k$ allows to extract $k$ bits at once. Extracting the least significant base $b$ digit means: $x\rightarrow \lfloor x/b \rfloor$ and $\textrm{mod}(x,b)$ goes to the bitstream.

Observe that taking interval in form ($l\in\mathbb{N}$):
\be I:=\{l,l+1,..,lb-1\}\ee
for any $x\in\mathbb{N}$ we have exactly one of three cases:
\begin{itemize}
  \item $x\in I$ or
  \item $x>lb-1$, then $\exists!_{k\in\mathbb{N}}\ \lfloor x/b^k\rfloor\in I$ or
  \item $x<l$, then $\forall_{(d_i)_i\in \{0,..,b-1\}^\mathbb{N}}\
  \exists!_{k\in\mathbb{N}}\ xb^k+d_1 b^{k-1}+..+d_k\in I$.
\end{itemize}

We will refer to this kind of interval as $b$-\emph{unique} as  eventually inserting ($x\to bx+d$) or removing ($x\to \lfloor x/b \rfloor$) some of the least significant digits of $x$, we will always finally get to $I$ in a unique way. Let us also define
\be I_s=\{x:C(s,x)\in I\}\qquad \textrm{so}\qquad I=\bigcup_s C(s,I_s) \ee
We can now define single steps for stream decoding: use $D$ function and then get the least significant digits from the steam until we get back to $I$, and encoding: send the least significant digits to the stream, until we can use the $C$ function:\\

\begin{tabular}{l|l}
  \textbf{Stream decoding}: & \textbf{Stream encoding}($s$):\\
\verb"{"$(s,x)=D(x)$\verb";"
&  \verb"{while("$x\notin I_s$\verb")"\\
\verb" use "$s$\verb";  "(e.g. to generate symbol)
&  \verb"   {put mod("$x,b$\verb") to output; "$x=\lfloor x/b\rfloor$\verb"}"\\
\verb" while("$x\notin I$\verb")"
& \verb" "$x=C(s,x)$\\
\verb"     "$x=xb+$\verb"'digit from input'" & \verb"}"\\
\verb"}"
\end{tabular}\\

Above functions are inverse of each other if only $I_s$ are also $b$-unique: there exists $\{l_s\}_{s\in\mathcal A}$ such that
\be I_s=\{l_s,...,l_s b-1\} \label{nec1} \ee
We need to ensure that this necessary condition is fulfilled, what is not automatically true as we will see in the example. More compact form of this condition will be found in Section \ref{necabs} and for larger alphabets we will directly enforce its fulfillment in Section \ref{ans}.\\

In practice we do not need to transfer these digits one by one like in the pseudocode above, what needs multiple steps of slow branching and is usually required in arithmetic coding. While using coding tables we can directly store also the number of bits to transfer and perform the whole transfer in a single unconditional operation, what is used in fast implementation of Yann Collet \cite{fse}. For $b=2$ and $l$ being a power of 2, the number of bits to transfer while decoding step is the number of bits in the buffer minus the current length of $x$ (implemented in modern processors). For encoding given symbol, the corresponding number of bits to transfer can take only one of two succeeding values.
\subsection{Example} \label{exs}
Taking uABS for $p=0.3$ as in Fig. \ref{intr}, let us transform it to stream version for $b=2$ (we transfer single least significant bits). Choosing $l=8$, we have  $x\in I=\{8,..,15\}$. We can see from this figure that $I_0=\{5,..,10\}$, $I_1=\{3,4\}$ are the ranges from which we get to $I$ after encoding correspondingly 0 or 1. These ranges are not $b$-unique, so for example if we would like to encode $s=1$ from $x=10$, while transferring the least significant bits we would first reduce it to $x=5$ and then to $x=2$, not getting into the $I_1$ range.

However, if we choose $l=9$, $I=\{9,..,17\}$, we get $I_0=\{6,..,11\}$ and $I_1=\{3,4,5\}$ which are $b$-unique. Here is the encoding process for $s=0,1$: first we transfer some of the least significant bits to the stream, until we reduce $x$ to $I_s$ range.  Then we use ABS formula (like in Fig. \ref{intr}):

\begin{center}\begin{tabular}{c|c|c|c|c|c|c|c|c|c|}
  $x\in I$ & 9 & 10 & 11 & 12 & 13 & 14 & 15 & 16 & 17 \\\hline
  bit transfer for $s=0$ & - & - & - & 0 & 1 & 0 & 1 & 0 & 1 \\
  reduced $x'\in I_0$ & 9&10&11&6&6&7&7&8&8 \\
  $\overline{C}(0,x)=C(0,x')$ & 14 & 15 & 17 & 9 & 9 & 11 & 11 & 12 & 12 \\  \hline
  bit transfer for $s=1$  & 1 & 0 & 1 & 0, 0 & 1, 0 & 0, 1 & 1, 1 & 0, 0 & 1, 0 \\
  reduced $x'\in I_1$  & 4 & 5 & 5 & 3 & 3 & 3 & 3 & 4 & 4 \\
  $\overline{C}(1,x)=C(1,x')$   & 13& 16&16 & 10 &10&10& 10 & 13 &13 \\  \hline
\end{tabular}\end{center}

Here is an example of evolution of (state, bit sequence) while using this table:
$$ (9,-)\xrightarrow{1} (13,1)\xrightarrow{0} (9,11) \xrightarrow {0} (14,11) \xrightarrow{1} (10,1101)\xrightarrow{0} (15,1101)\xrightarrow{1} (10,110111) ... \label{ex2}$$

The decoder will first use $D(x)$ to get symbol and reduced $x'$, then add the least significant bits from stream (in reversed order).\\

To find the expected number of bits/symbol used by such process, let us first find its stationary probability distribution assuming i.i.d. input source. Denoting by $\overline{C}(s,x)$ the state to which we go from state $x$ due to symbol $s$ (like in the table above), this stationary probability distribution have to fulfill:
\be \Pr(x)=\sum_{s,y:\ \overline{C}(s,y)=x} \Pr(y)\cdot p_s \label{station} \ee
equation, being the dominant eigenvector of corresponding stochastic matrix. Such numerically found distribution is written in Fig. \ref{auto}. Here is approximated distribution for the currently considered automaton and comparison with close $\Pr(x)\propto 1/x$ distribution, what will be motivated in Section \ref{statio}:\\

\begin{tabular}{c|c|c|c|c|c|c|c|c|c|}
   $x$ & 9 & 10 & 11 & 12 & 13 & 14 & 15 & 16 & 17 \\ \hline
  $\Pr(x)$   & 0.1534 & 0.1240 & 0.1360 & 0.1212 & 0.0980 & 0.1074 & 0.0868 & 0.0780 & 0.0952 \\
  $1.3856/x$ & 0.1540 & 0.1386 & 0.1260 & 0.1155 & 0.1066 & 0.0990 & 0.0924 & 0.0866 & 0.0815\\
\end{tabular}\\

We can now find the expected number of bits/symbol used by this automaton by summing the number of bits used in encoding table:
$$\left(\sum_{x=12}^{17} \Pr(x)\right)p_0+\left(\Pr(9)+\Pr(10)+\Pr(11)+2\sum_{x=12}^{17} \Pr(x)\right)p_1 \approx 0.88658 \textrm{ bits/symbol}$$
For comparison, Shannon entropy is $-p\lg(p)-(1-p)\lg(1-p)\approx 0.88129$ bits/symbol, so this automaton uses about $\Delta H\approx 0.00529$ bits/symbol more than required.\\

The found encoding table fully determines the encoding process and analogously we can generate the table for decoding process, defining automaton like in Fig. \ref{auto}.

Observe that having a few such encoders operating on the same range $I$, we can use them consecutively in some order - if decoder will use them in reversed order, it will still retrieve the encoded message. Such combination of multiple different encoders can be useful for example when probability distribution (or variant or even alphabet size) varies depending on context, or for cryptographic purposes.
\subsection{Necessary condition and remarks for stream formulas} \label{necabs}
For unique decoding we need that all $I_s$ are $b$-unique, what is not always true as we have seen in the example above. In the next section we will enforce it by construction for larger alphabet. We will now check when the uABS, rABS and rANS formulas fulfill this condition.

In the uABS case we need to ensure that both $I_0$ and $I_1$ are $b$-absorbing: denoting $I_s=\{l_s,..,u_s\}$, we need to check that
$$u_s=bl_s-1 \quad\qquad \textrm{for }s=0,1$$
We have $l_s=|\{x<l:\overline{s}(x)=s\}|$, $u_s=|\{x<bl:\overline{s}(x)=s\}|-1$, so $\sum_s l_s=l$, $\sum_s u_s=bl-2$. It means that fulfilling one of these condition implies the second one.

Let us check when $u_1=bl_1-1$. We have $l_1=\lceil lp\rceil$, $u_1=\lceil blp\rceil -1$, so the condition is:\\
\textbf{Condition:} Stream uABS can be used when
 \be b \lceil lp\rceil=\lceil blp\rceil.\ee
The basic situation this condition is fulfilled is when $lp\in\mathbb{N}$, what means that $p$ is defined with $1/l$ precision.\\

For rABS and rANS, for simplicity let us assume that $m$ divides $l$: $l=km$ (there can be also other possibilities). In this case, $kl_s$ is the first appearance of symbol $s$ in $I$, $bkl_s-1$ is the last one as required.\\
\textbf{Condition:} If $m$ divides $l$, rABS and rANS can be used in stream version.\\

There are two basic possibilities for using coding formulas (\cite{mah} implementations):
\begin{itemize}
  \item{use them directly for example in 32 bit arithmetics (fpaqc). As $l>2^{20}$, inaccuracy becomes negligible, we can use large $b$ to extract multiple bits at once. However, the multiplication can make it a bit slower,}
  \item{store behavior on some chosen range (fpaqb) - it is less accurate, but can be a bit faster and leaves freedom to choose exact encoding - we will explore this possibility with tANS.}
\end{itemize}
If the probability varies, in the former case we just use the formula for current $p$, while in the latter case we should have prepared tables for different probability distributions (e.g. quantized) we could use for approximation. Such varying of probability distribution is allowed as long $I$ remains fixed.
\subsection{Analysis of a single step} \label{stepan}
Let us now take a closer look at a single step of stream encoding and decoding. Observe that the number of digits to transfer to get from $x$ to $I_s=\{l_s,..,bl_s-1\}$ is  $k=\lfloor \log_b(x/l_s)\rfloor$, so  the (new symbol, digit sequence) while stream coding is:
\be x \xrightarrow{s}  \left(C(s, \lfloor x/b^k \rfloor),\ \textrm{mod}(x,b^k)\right) \qquad\qquad \textrm{where}\quad  k=\lfloor \log_b(x/l_s)\rfloor \ee
this $\textrm{mod}(x,b^k)=x-b^k\lfloor x/b^k \rfloor $ is the information lost from $x$ while digit transfer - this value is being sent to the stream in base $b$ numeral system. The original algorithm produces these digits in reversed order: from less to more significant.

\begin{figure}[b!]
    \centering
        \includegraphics{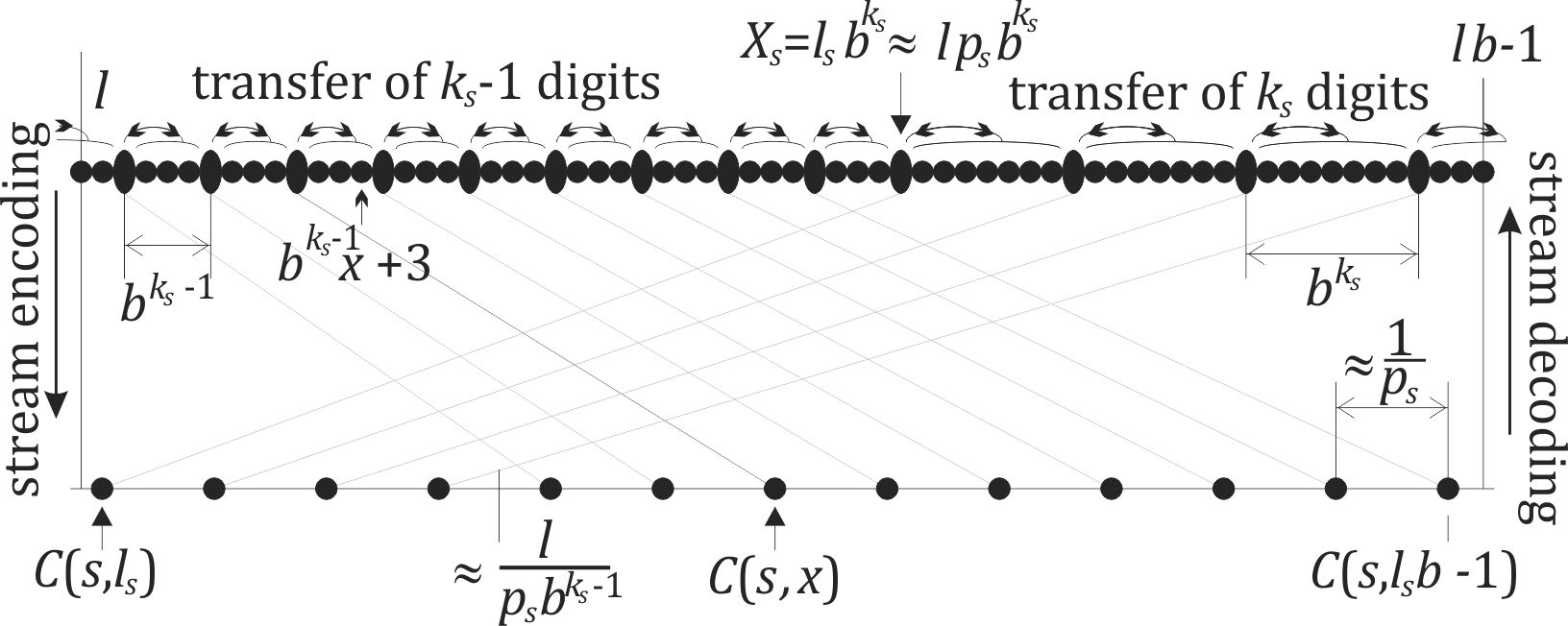}
        \caption{Example of stream coding/decoding step for $b=2,\ k_s=3,\ l_s=13,\quad$  $l=9\cdot 4+3\cdot 8+6=66,\ p_s=13/66,\ x=19,\ b^{k_s-1}x+3=79=66+2+2\cdot 4+3$.}    \label{step}
\end{figure}

Observe that for given symbol $s$, the number of digits to transfer can obtain one of two values: $k_s-1$ or $k_s$ for $k_s:=\lfloor \log_b(x/l_s)\rfloor+1$. The smallest $x$ requiring to transfer $k_s$ digits is $X_s:=lb^{k_s}$. So finally the number of bits to transfer while encoding is $k_s-1$ for $x\in\{l,..,X_s-1\}$ and $k_s$ for $x\in \{X_s,..,lb-1\}$, like in Fig. \ref{step}.\\

While stream decoding, the current state $x$ defines currently produced symbol $s$. In example in Fig. \ref{auto}, the state also defined the number of digits to take: 1 for $x=4$, 2 for $x=5$ and 0 for $x=6,7$.

However, in example from Section \ref{exs}, state $x=13$ is an exception: it is decoded to $s=1$ and $x'=4$. Now if the first digit is $1$, the state became $x=2\cdot4+1=9\in I$. But if the first digit is $0$, it became $2\cdot 4=8\notin I$ and so we need to take another digit, finally getting to $x=16$ or $17$. We can also see such situation in Fig. \ref{step}.

This issue - that decoding from a given state may require $k_s-1$ or $k_s$ digits to transfer, means that for some $x\in \mathbb{N}$ we have: $bx<l$, while $bx+b-1\geq l$. It is possible only if $b$ does not divide $l$.

To summarize: if $b$ divides $l$, decoding requires to transfer a number of digits which is fixed for every state $x$. So we can treat these digits $(\textrm{mod}(x,b^k))$ as blocks and store them in forward or in backward order. However, if $b$ does not divide $l$, the state itself sometimes does not determine the number of digits: we should first take $k_s-1$ first digits, and if we are still not in $I$, add one more. We see that in this case, it is essential that digits are in the proper order (reversed): from the least to the most significant.
\subsection{Stationary probability distribution of states}\label{statio}
In Section \ref{exs} there was mentioned finding the stationary probability distribution of states used while encoding (and so also while decoding) - fulfilling (\ref{station}) condition. We will now motivate that this distribution is approximately $\Pr(x)\propto 1/x$.\\

\begin{figure}[b!]
    \centering
        \includegraphics{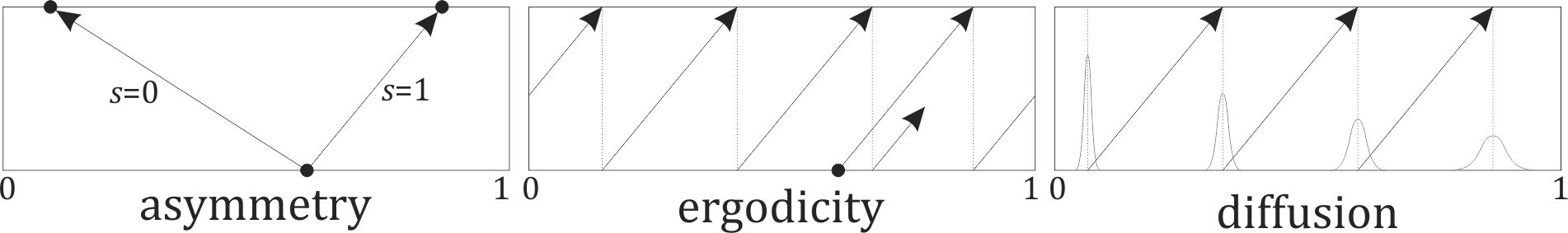}\
        \caption{Three sources of chaotic behavior of $y=\log_b (x/l)$, leading to nearly uniform distribution on the $[0,1]$ range.}
        \label{chaos}
\end{figure}

Let us transform $I$ into $[0,1)$ range (informational content):
\be y=\log_b \left(x/l\right) \in [0,1)\ee
Encoding symbol of probability $p_s$ increases $y$ by approximately $\log_b(1/p_s)$. However, to remain in the range, sometimes we need to transfer some number of digits before. Each digit transfer reduces $x$ to $\lfloor x/d\rfloor$ and so $y$ by approximately 1. Finally a single step is:
$$y \xrightarrow{s}\ \approx \{ y+\log_b(1/p_s)\}$$
where $\{a\}=a-\lfloor a \rfloor$ denotes the fractional part.

While neighboring values in arithmetic coding remain close to each other during addition of succeeding symbols (ranges are further compressed), here we have much more chaotic behavior, making it more appropriate for cryptographic applications. There are three sources of its chaosity, visualized in Fig. \ref{chaos}:
\begin{itemize}
  \item \emph{asymmetry}: different symbols can have different probability and so different shift,
  \item \emph{ergodicity}: $\log_b(1/p_s)$ is usually irrational, so even a single symbol should lead to uniform covering of the range,
  \item \emph{diffusion}: $\log_b(1/p_s)$ only approximates the exact shift and this approximation varies with position - leading to pseudorandom diffusion around the expected position.
\end{itemize}

These reasons suggest that probability distribution while encoding is nearly uniform for $y$ variable: $\Pr(y\leq a)\approx a\in [0,1]$ , what is confirmed by numerical simulations. Transforming it back to $x$ variable, probability of using $x\in I$ state is approximately proportional to $1/x$
\be \Pr(x)\propto^\sim 1/x\ee

It means that the informational content of being in state $x$ is $\lg(1/ \Pr(x))\approx \lg(x) +\textrm{const}$, what was the initial concept behind ANS.
\subsection{Bound for $\Delta H$}
We will now find some upper bound for the distance from Shannon entropy $\Delta H$. As finding the exact stationary probability is a complex task and this distribution can be disturbed by eventual correlations, we will use the absolute bound for inaccuracy: $|\epsilon_s(x)|=|x_s/x-p_s|\leq 1/x$ for uABS ($m/x$ for rANS) to find an independent bound.
\begin{figure}[t!]
    \centering
        \includegraphics{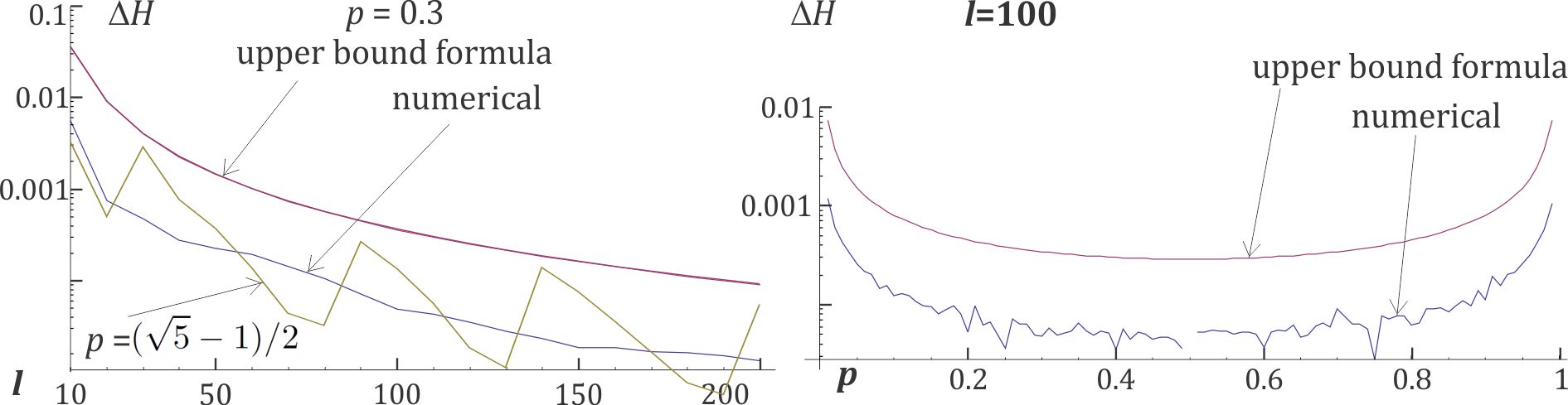}\
        \caption{Left: $\Delta H$ for $p=0.3$ uABS and different $l$ being multiplicities of 10. The red line is formula (\ref{bound}). Additionally, the irregular yellow line is situation in these $l$ when we approximate some irrational number using fractions with denominator $l$. Right: situation for fixed $l=100$ and $p=i/100$ for $i=1,2,..,99$. Discontinuity for $i=50$ is because $\Delta H=0$ there.}
        \label{abssim}
\end{figure}

Generally, assuming that the probability distribution is indeed $\{p_s\}$, but we pay $\lg(1/q_s)$ bits per symbol $s$, the cost of inaccuracy is the Kullback-Leiber distance:
$$\Delta H=\sum_s p_s \lg(1/q_s)- \sum_s p_s \lg(1/p_s)=-\sum_s p_s\lg\left(1-\left(1-\frac{q_s}{p_s}\right)\right)=$$
$$ =\sum_s \frac{p_s}{\ln(2)}\left(\left(1-\frac{q_s}{p_s}\right)+\frac{1}{2}\left(1-\frac{q_s}{p_s}\right)^2+O\left(\left(1-\frac{q_s}{p_s}\right)^3\right)\right)
\approx \sum_s \frac{(p_s-q_s)^2}{p_s\ln(4)} $$

In our case we use $\lg(x/x_s)$ bits to encode symbol $s$ from state $x_s$, so $(p_s-q_s)^2$ corresponds to $|\epsilon_s(x)|^2<1/l^2$ for $x\in I=\{l,..,lb-1\}$. The above sum should also average over the encoder state, but as we are using a general upper bound for $\epsilon$, we can bound the expected capacity loss of uABS on $I=\{l,..,bl-1\}$ range to at most:
\be \Delta H \leq \frac{1}{l^2\ln(4)}\sum_{s=0,1} \frac{1}{p_s}+O(l^{-3}) \ \textrm{bits/symbol} \label{bound} \ee
From Fig. \ref{abssim} we see that it is a rough bound, but it reflects well the general behavior. For rANS this bound should be multiplied by $m$.
\subsection{Initial state and direction of encoding/decoding} \label{direction}
Encoding and decoding in discussed methods are in reversed direction - what is perfect if we need a stack for symbols, but often may be an inconvenience. One issue is the need for storing the final state of encoding, which will be required to start decoding - the cost is usually negligible: a few bits of information for every data block.

However, if this cost is essential, we can avoid it be encoding information also in this initial state. For this purpose, instead of starting for example with $x=l$, we can:
\begin{itemize}
  \item start with $x$ having a single or more symbols directly stored in its least significant bits, like $x=l+s$,
  \item while using formulas, start with $x=1$ and encode succeeding symbols until getting close to $I$, then use the obtained $x$ as the initial state (can be a bit smaller than $l$),
  \item we can do analogously while using tables, but it requires more memory to define them also for $\{2,..,l-1\}$ range.
\end{itemize}

\begin{figure}[b!]
    \centering
        \includegraphics{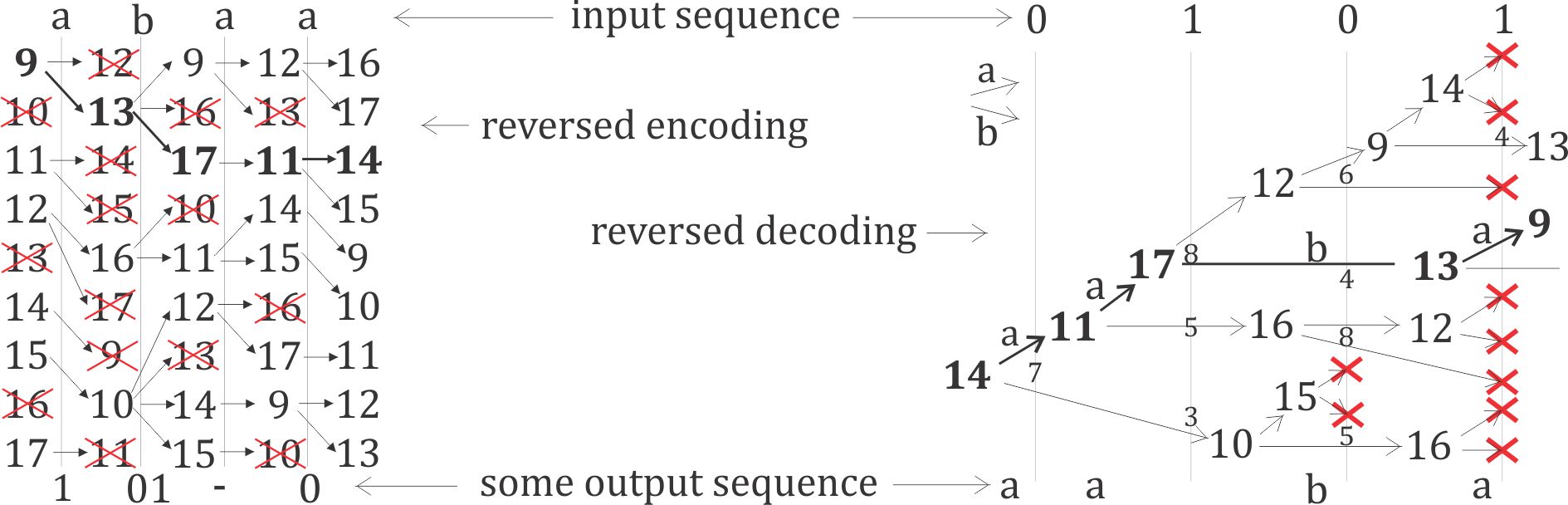}\
        \caption{Example of encoding (left) and decoding (right) in reversed direction for encoder from Section \ref{exs} - it is much more costly, but still possible. We find some output sequence, such that later correspondingly decoding or encoding in the same direction will produce the input sequence. While reversed encoding we can start with all possible states, then in every step remove those not corresponding to given symbol and make a step for every possible digits. While reversed decoding, here from a single state, we try to encode all possible symbols from given state (arrow up: $a$, arrow down: $b$) and check if the input bits agree with the least significant bits. The written small numbers are after removing this least significant bit. Finally, in both cases we choose a single path. There is always such a path as we could perform this encoding/decoding in standard direction starting with any state.}
        \label{reversed}
\end{figure}

Another issue is when probabilities depend on already processed data - in this case we can encode the data in backward direction, using information available while later decoding in forward direction. For example for Markov source of $s_1..s_m$ symbols, we would encode them in backward direction: from $m$-th to the 1-st, but the probability used for $s_k$ would depend on $s_{k-1}$. For adaptive encoding, we would need to process the whole data block in forward direction, assigning to each position probability used while decoding, then encode in backward direction using theses probabilities. Thanks of that, we can later use standard Markov or adaptive forward decoding.\\

However, if it is really necessary to encode and decode in the same direction, it is possible to change direction of encoding or decoding, but it is much more costly. We can see example of a few steps of such process in Fig. \ref{reversed}.
\section{Tabled asymmetric numeral systems (tANS)} \label{ans}
We will now focus on probably the most significant advantage of ANS approach over AC - thanks to smaller states, storing the entire behavior for large alphabet in relatively small coding table, getting usually faster coding than for Huffman, with accuracy/compression rate like in arithmetic coding. Instead of splitting a large symbol into binary choices, we can process it in a simple unconditional loop with a single table use and a few binary operations. Also instead of slow bit by bit AC renormalization, all bits can now be transferred at once. The cost is storing the tables: let say 1-16kB for 256 size alphabet. We can store many of them for various contexts/probability distributions, alphabet sizes and, as long as they use the same state space $I$, just switch between them.\\

Instead of trying to find large alphabet analogues of uABS formulas, we will directly generate tables used for encoding/decoding, like in example from Section \ref{exs}. There is a large freedom of choice for such a table, as every symbol distribution corresponds to a different coding. The distance from the Shannon limit depends on how well we approximate the $x_s/x\approx p_s$ relation. We will now focus on doing it in a very accurate way to get very close to capacity limit. However, this process can be for example disturbed in pseudorandom way using a cryptographic key to additionally encrypt the message.
\subsection{Precise initialization algorithm}
Assume we have given $l$, $b$ and probability distribution of $n$ symbols: $0 < p_1,..,p_n <1$, $\sum_s p_s=1$. For simplicity, let us assume for now that probabilities are defined with $1/l$ accuracy:
\be l_s:=lp_s \in \mathbb{N} \label{approx}\ee
For general probabilities, in Section \ref{qabs} we will see that we can "tune" the coder by shifting symbol appearances to make it optimal for slightly different probability distributions. Shifting symbol appearances right (to larger $x$) generally corresponds to increasing its informational content and so decreasing probability ($\approx x_s/s$).
  
\begin{figure}[b!]
    \centering
        \includegraphics{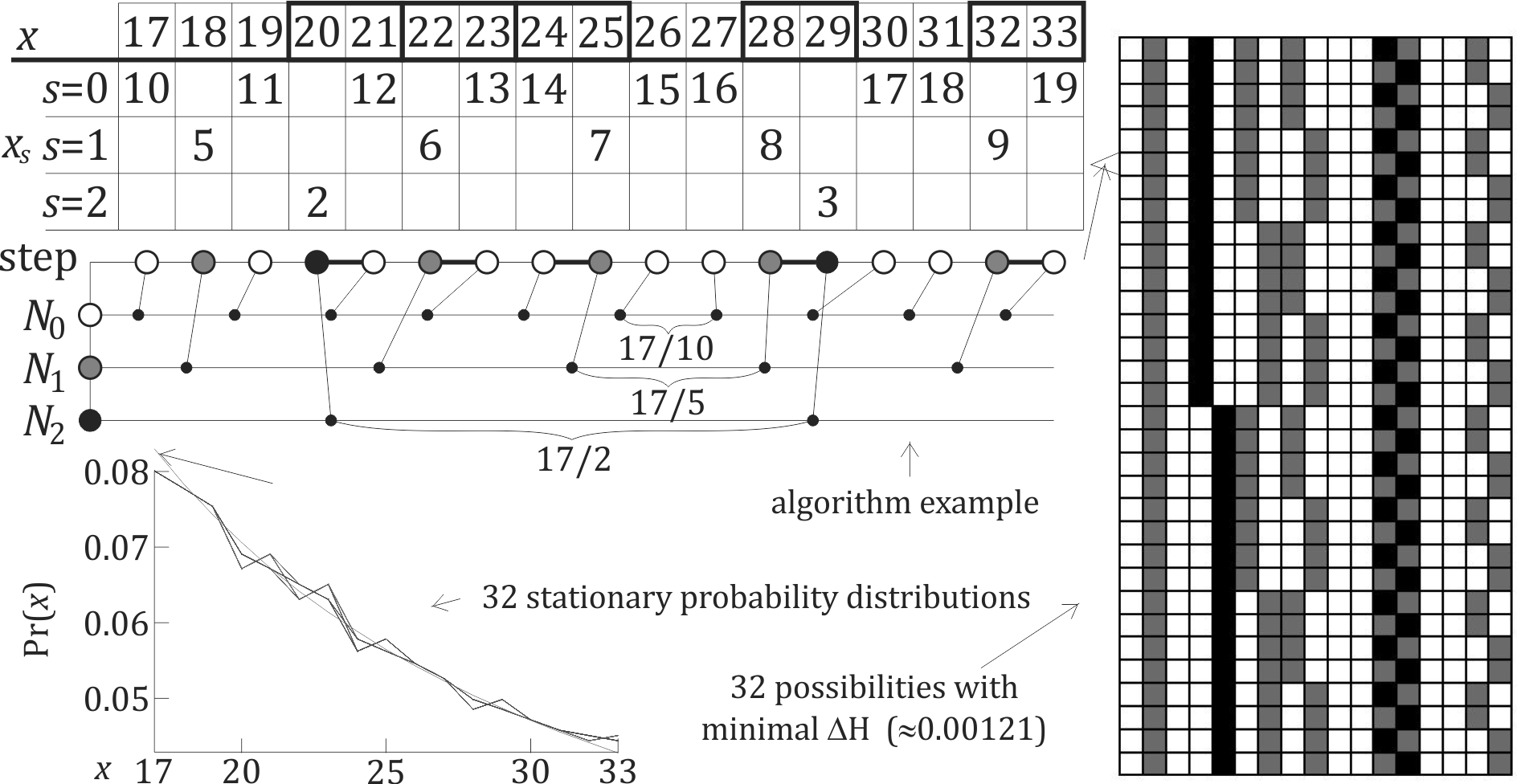}\
        \caption{Left from top: precise algorithm initialization for $(10,5,2)/17$ probability distribution, corresponding encoding table and stationary probability distributions. Right: among ${17 \choose 2,5}$ possibilities to choose symbol distribution here, the minimal $\Delta H$ turns out to be obtained by 32 of them - presented in graphical form. The sixth from the top is the one generated by our algorithm. These 32 possibilities turn out to differ by switching two neighboring positions in 5 locations ($2^5=32$) - these pairs are marked by thick lines. We can see that these 5 positions correspond to the largest deviations from $\propto 1/x$ expected behavior of stationary probability distribution.}
        \label{algor}
\end{figure}

The encoding is defined by distributing symbols in nearly uniform way on the $I=\{l,..,bl-1\}$ range: $(b-1)l_s$ appearances of symbol $s$. Then for every symbol $s$ we enumerate its appearances by succeeding numbers from $I_s=\{l_s,..,bl_s-1\}$ range, like in Fig. \ref{algor}. \\

Unfortunately, finding the optimal symbol distribution is not an easy task. We can do it by checking all symbol distributions and finding $\Delta H$ for each of them - for example for $l_1=10$, $l_2=5$, $l_3=2$ there are ${17 \choose 2,5}=408408$ ways to distribute the symbols and each of them corresponds to a differed coding. While testing all these possibilities it turns out that the minimal value of $\Delta H$ among them: $\Delta H\approx 0.00121$ bits/symbol, is obtained by 32 different distributions - they are presented in the right hand side part of Fig. \ref{algor}. They can be obtained by switching pairs of neighboring nodes on some 5 positions (marked with thick lines) - bit sequences produced by them differ by corresponding bit flips.

We will now introduce and analyze a heuristic algorithm which directly chooses a symbol distribution in nearly uniform way. Example of its application is presented in Fig. \ref{algor}. In this case it finds one of the optimal distributions, but it is not always true - sometimes there are a bit better distributions than the generated one. So if the capacity is a top priority while designing a coder, it can be reasonable to check all symbol distributions and generally we should search for better distributing algorithms.\\

To formulate the heuristic algorithm, let us first define:
\be N_s:=\left\{\frac{1}{2p_s}+\frac{i}{p_s}:\ i\in \mathbb{N}\right\} \ee

These $n$ sets are uniformly distributed with required densities, but they get out of $\mathbb{N}$ set - to define ANS we need to shift them there, like in Fig. \ref{algor}. Specifically, to choose a symbol for the succeeding position, we can take the smallest not used element from $N_1,..,N_n$ sets. This simple algorithm requires a priority queue to retrieve the smallest one among currently considered $n$ values - if using a heap for this purpose, the computational cost grows like $\lg(n)$ (instead of linearly for greedy search). Beside initialization, this queue needs two instructions: let \verb"put("$(v,s)$\verb")" insert pair $(v,s)$ with value $v\in N_s$ pointing the expected succeeding position for symbol $s$. The second instruction: \verb"getmin" removes and returns pair which is the smallest for $(v,s)\leq(v',s')\Leftrightarrow v\leq v'$ relation.\\

Finally the algorithm is:\\
 \textbf{Precise initialization} of decoding or encoding function/table

\noindent \verb"For "$s=0$\verb" to " $n-1$ \verb" do {put("
$(0.5/p_s,s)$ \verb"); "$x_s=l_s$\verb"};"\\
\verb"For "$x=l$\verb" to "$bl-1$\verb" do"\\
\verb"  {"$(v,s)$\verb"=getmin; put("$(v+1/p_s,s)$\verb");"\\
\verb"   "$D[x]=(s,x_s)$ or $C[s,x_s]=x$ \\
\verb"   "$x_s$\verb"++}"\\

There has remained a question which symbol should be chosen if two of them have the same position. Experiments suggest to choose the least probable symbol among them first, like in example in Fig. \ref{algor} (it can be realized for example by adding some small values to $N_s$). However, there can be found examples when opposite approach: the most probable first, brings a bit smaller $\Delta H$.
\subsection{Inaccuracy bound}
Let us now find some bound for inaccuracy $\epsilon_s(x)=x_s/x-p_s$ for this algorithm to get an upper bound for $\Delta H$. Observe that the symbol sequence it produces has period $l$, so for simplicity we can imagine that it starts with $x=0$ instead of $x=l$.

From definition, $\# (N_s \cap [0,x-1/(2p_s)])=\lfloor x p_s \rfloor$. As $\lfloor x p_s\rfloor \leq x p_s\leq \lfloor x p_s+1\rfloor$ and $\sum_s p_s=1$, we have also $\sum_s \lfloor xp_s \rfloor \leq x\leq \sum_s \lfloor xp_s+1 \rfloor$. Defining $\overline{p}=\min_s p_s$, we can check that in $[0,x-1/(2\overline{p})]$ range there is at most $x$ symbols in all $N_s$ and in $[0,x+1/(2\overline{p})]$ range there is at least $x$ symbols:
$$  \sum_s \# \left(N_s \cap \left[0,x-\frac{1}{2\overline{p}}\right]\right)\leq  \sum_s \# \left(N_s \cap \left[0,x-\frac{1}{2p_s}\right]\right)= \sum_s \lfloor xp_s \rfloor \leq x $$

\begin{figure}[b!]
    \centering
        \includegraphics{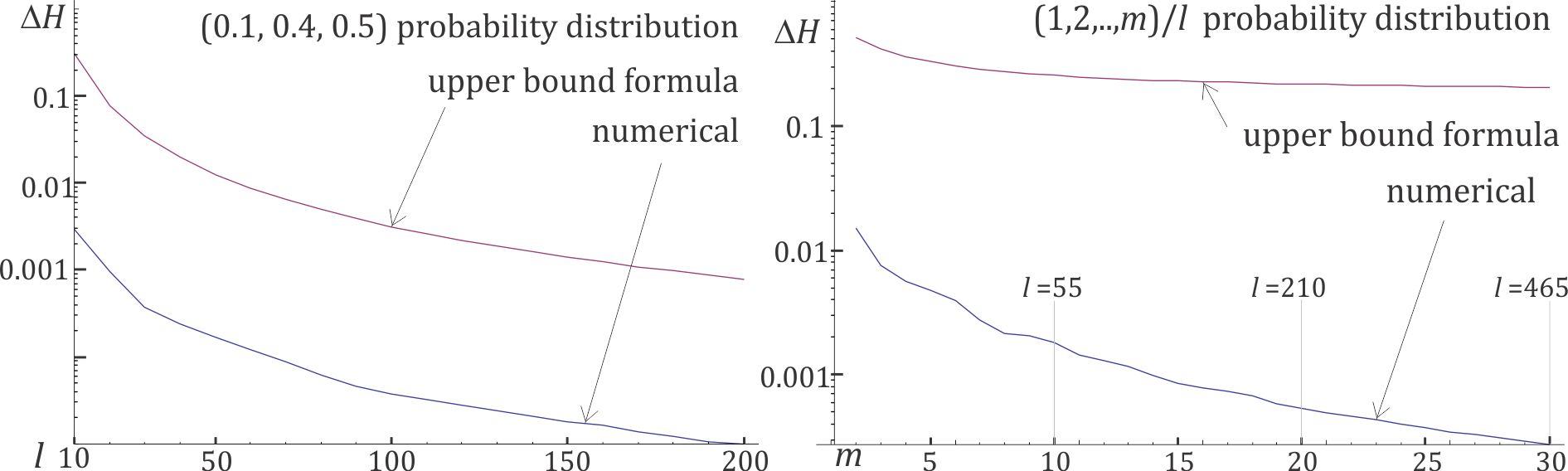}\
        \caption{Left: numerical values and comparison with the (\ref{bound1}) formula for $(0.1,0.4,0.5)$ probability distribution and $l$ being a multiplicity of 10. Right: comparison for larger alphabet: of size $m$. The probability distribution is $(1,2,..,m)/l$, where $l=\sum_{i=1}^m i=m(m+1)/2$.}
        \label{anssim}
\end{figure}

$$x\leq \sum_s \lfloor xp_s+1 \rfloor = \sum_s \# \left(N_s \cap \left[0,x+\frac{1}{2p_s}\right]\right) \leq \sum_s \# \left(N_s \cap \left[0,x+\frac{1}{2\overline{p}}\right]\right) $$
So $x$-th symbol found by the algorithm had to be chosen from $\left[x-1/(2\overline{p}),x+1/(2\overline{p})\right]\cap\bigcup_s N_s$. From definition, the $x_s$-th value of $N_s$ is $1/(2p_s)+x_s/p_s$. If this value corresponds to $x$, it had to be in the $\left[x-1/(2\overline{p}),x+1/(2\overline{p})\right]$ range:
$$\left| \frac{1}{2p_s}+\frac{x_s}{p_s} -x \right|\leq \frac{1}{2\overline{p}}\qquad\qquad \Rightarrow\qquad\qquad \left|\frac{1}{2x} + \frac{x_s}{x} - p_s\right|\leq \frac{p_s}{2x\overline{p}}$$
\be \left|\epsilon_s(x)\right|=\left|\frac{x_s}{x}-p_s\right|  \leq \left(\frac{p_s}{2\overline{p}}+\frac{1}{2}\right)/x\ee
Finally in analogy to (\ref{bound}) we get:
\be \Delta H \leq \frac{1}{l^2\ln(4)}\sum_s\frac{1}{p_s} \left(\frac{p_s}{2\min_{s'} p_{s'}}+\frac{1}{2}\right)^2+O(l^{-3}) \ \textrm{bits/symbol} \label{bound1} \ee

\begin{figure}[b!]
    \centering
        \includegraphics{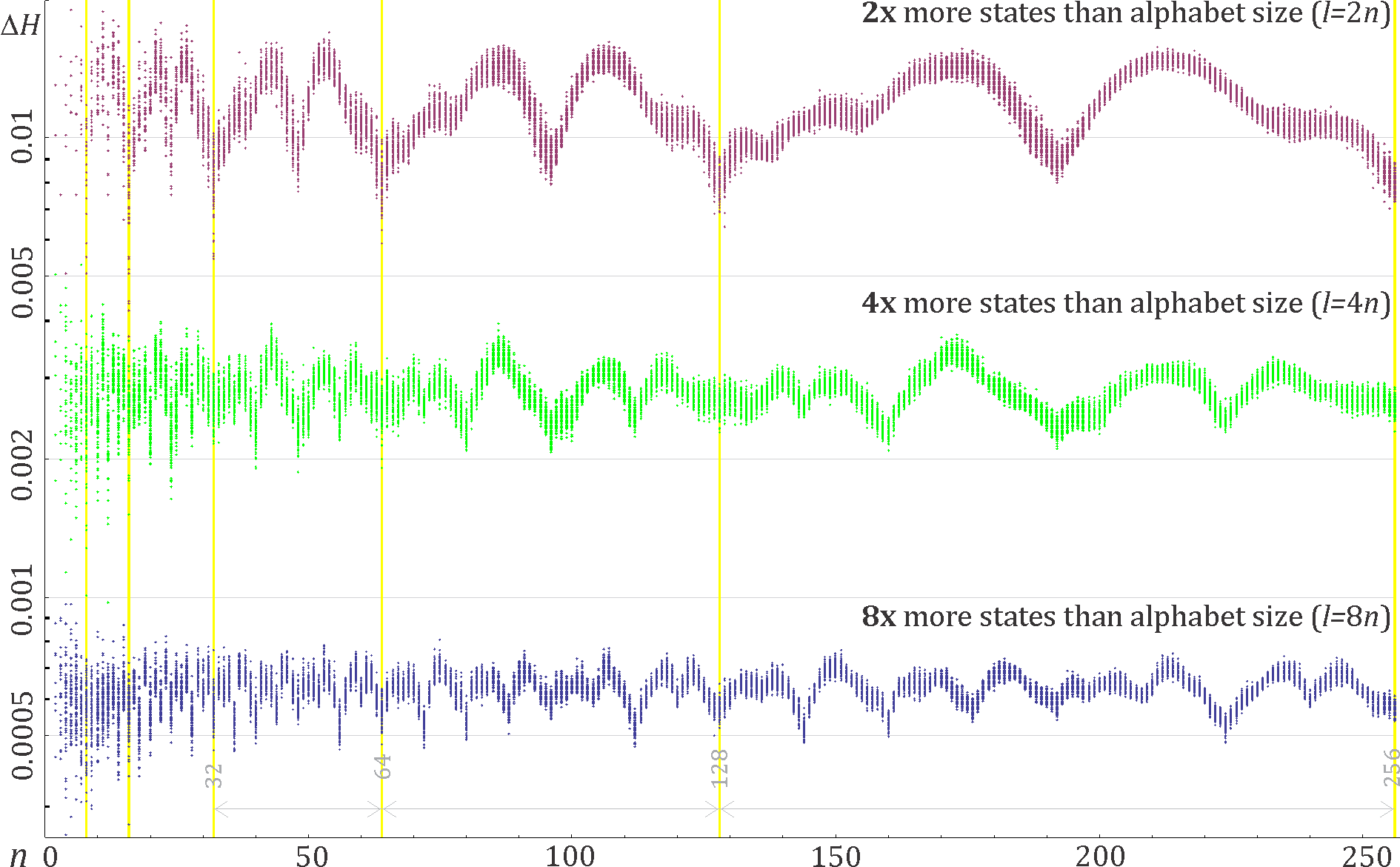}\
        \caption{$\Delta H$ for different alphabet size. The three graphs correspond to using the number of states ($l$, $b=2$) being 2, 4 and 8 times larger than alphabet size ($n=2,..,256$). For every $3\cdot 255$ case there was taken 100 random distributions generated by procedure: start with $l_s=1$ for all symbols, then repeat $l-n$ times: increase $l_s$ for a random $s$ by 1. The yellow lines mark powers of 2 - there is some self-similarity relation there, probably caused by using $b=2$.}
        \label{tans}
\end{figure}

From Fig. \ref{anssim} we can see that it is a very rough bound. Figure \ref{tans} shows results of simulations for practical application - tANS behavior for random probability distributions while using the number of states being a small multiplicity of alphabet size. While the behavior is quite complex, general suggestion is that using $k$ more states than alphabet size, $\Delta H\approx 0.5/k^2$. So if we want to work in $\Delta H\approx 0.01$ bits/symbol regime, we should use $k=2$ or 4, for $\Delta H\approx 0.001$ bits/symbol it should be rather 8 or 16.

These graphs do not take into consideration the fact that $l p_s$ usually are not natural numbers - that there is some additional approximation needed here. However, from Fig. \ref{abssim} we can see that such approximation can work both ways. We will now see that additionally we can tune symbol distribution to design coders with properly shifted optimal probability distribution.
\subsection{Quasi ABS (qABS) and tuning} \label{qabs}

\begin{figure}[b!]
    \centering
        \includegraphics{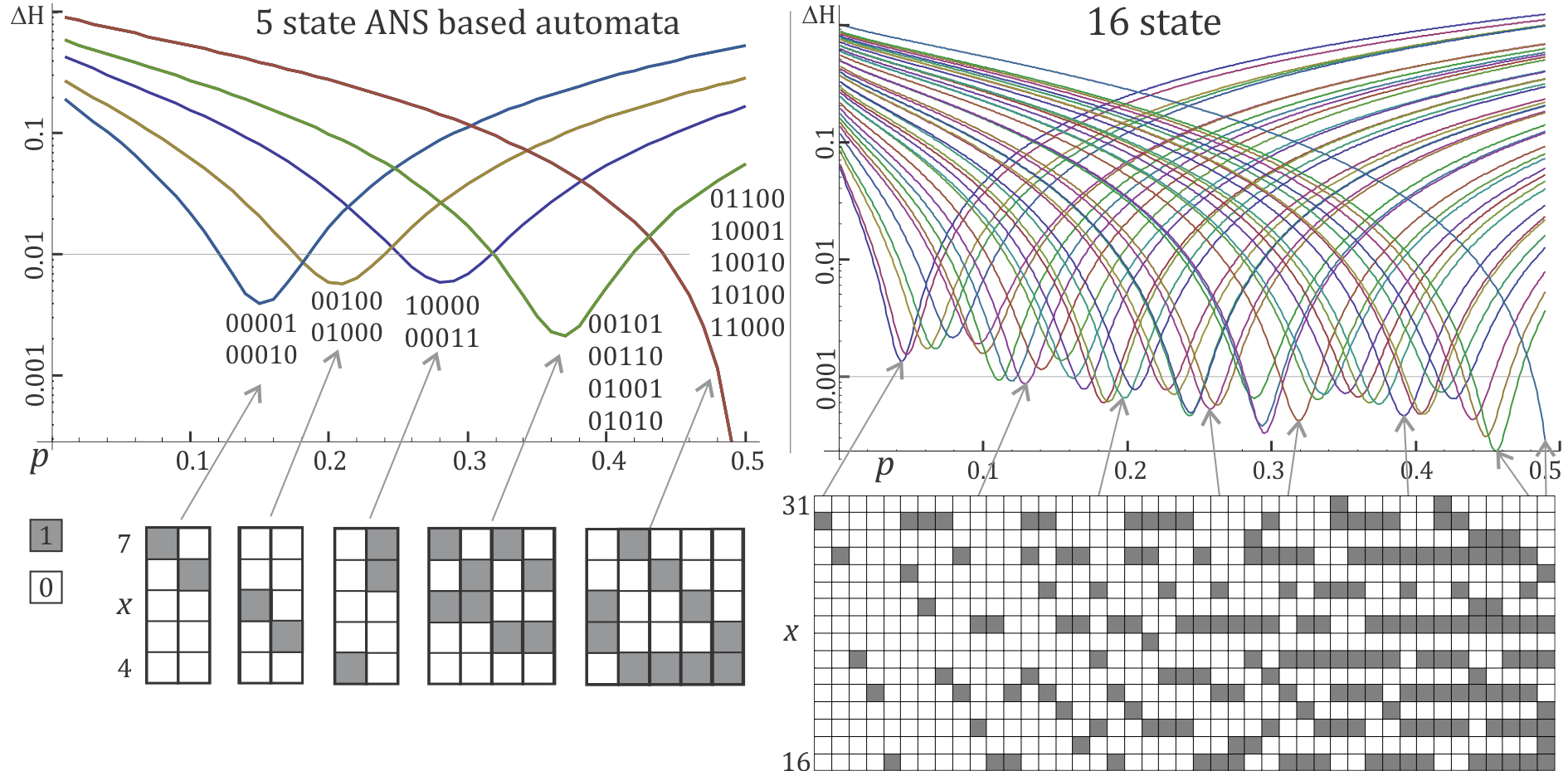}\
        \caption{Left: symbol distributions and $\Delta H$ for different symbol probabilities and all 4 state automata with more "0" symbol appearances than "1". Right: analogously for 16 states, but this time presenting only a single representant for each curve which is the lowest one for some $p$ - there are 43 of them.}
        \label{qans}
\end{figure}
Let us now change the way of thinking for a moment: instead of focusing on symbol distribution for probabilities approximated as $l_s/l$, let us test $\Delta H$ for all probability distributions and different fixed coders (symbol distributions) - some result can be seen in Figure \ref{qans}. Surprisingly, for 5 state automata, all symbol distributions turn out to be useful in some narrow range of probabilities - choosing between such 5 automata, we can work in $\Delta H\approx 0.01$ bits/symbol regime. For $\Delta H\approx 0.001$ bits/symbol, we should choose approximately 20 of shown 43 automata for 16 states - they have deeper single $\Delta H$ minimum, but also narrower.

This kind of approach is very similar to Howard-Vitter quasi arithmetic coding - Example 10 from \cite{quasi} presents analogous switching for 6 state automata. Calculating $\Delta H$ for i.i.d. static sources of different probabilities, it turns out that we obtain exactly the same graph as the 5 state case from Fig. \ref{qans} - this time ANS requires 1 state less. This agreement and further comparison of both approaches requires further investigation. However, the fact that ANS uses a single number state, while AC needs 2 number state for similar precision, suggests that the advantage of using ANS will grow.

Symbol distributions in Fig. \ref{qans} suggest that shifting symbol appearances toward lower $x$ generally shifts the optimal probability right. It can be understood that lowering $x$ here means increase of $x_s/x$, which is the probability we approximate. We could try to correspondingly modify the precise initialization this way: tune it when $p_s\neq l_s/l$, for example modifying the initial values of $N_s$, like using real $1/(2p_s)$ there instead of approximated by $l_s/l$. However, the general behavior is very complex and understanding it requires further work. For specific purposes we can find optimized symbol distributions, for example by exhaustive search like in Fig. \ref{algor}, and store them inside the compressor.\\

The fact that (000..01) symbol distribution corresponds to coder with the lowest optimal probability (left-most curves in Fig. \ref{qans}), suggests how to build cheap and simple coders for difficult to optimally handle situation: when $p\approx 0$. Automaton for this kind of distribution is: the more probable symbol ($1-p$) increases state by one ($X++$), until getting to the $2l-2$ state, from which (and state $2l-1$) it goes to the first state ($l$), producing a single bit. The less probable symbol ($p$) always gets to the last state ($2l-1$), producing all but the most significant bit of the current state. It produces approximately $\lg(l)$ bits, plus 1 for the succeeding step. Comparing this number to $\lg(1/p)$, we obtain that we should use $l\approx 0.5/p$.

\subsection{Combining tANS with encryption}
We will now briefly discuss using tANS to simultaneously encrypt the data or as a part of a cryptosystem. While standard cryptography usually operates on constant length bit blocks, encryption based on entropy coder has advantage of using blocks which lengths vary in a pseudorandom way. The inability to directly divide the bit sequence into blocks used in single steps makes cryptoanalysis much more difficult.

As it was mentioned in Section \ref{statio}, the behavior of the ANS state is chaotic - if someone does not know the exact decoding tables, he would quickly loose the information about the current state. While decoding from some two neighboring states, they often correspond to a different symbol and so their further behavior will be very different (asymmetry). In comparison, in arithmetic coding nearby values are compressed further - remain nearby.

The crucial property making ANS perfect for cryptographic applications is that, in contrast to arithmetic coding, it has huge freedom of choice for the exact coding - every symbol distribution defines a different encoding. We can use a pseudorandom number generator (PRNG) initialized with a cryptographic key to choose the exact distribution, for example by disturbing the precise initialization algorithm. One way could be instead of choosing the pair with the smallest $v$ by \verb"getmin" operation, use the PRNG to choose between $s$ which would be originally chosen and the second one in this order. This way we would get a bit worse accuracy and so $\Delta H$, but we have $2^{l(b-1)}$ different possibilities among which we choose the coding accordingly to the cryptographic key.

This philosophy of encryption uses the cryptographic key to generate a table which will be later used for coding. Observe that if we would make the generation of this table more computationally demanding, such approach would be more resistant to brute force attacks. Specifically, in standard cryptography the attacker can just start decryption with succeeding key to test if given key is the correct one. If we enforce some computationally demanding task to generate decoding tables (requiring for example 1ms of calculations), while decoding itself can be faster as we just use the table, testing large number of cryptographic keys becomes much more computationally demanding - this encryption philosophy can be made more resistant to unavoidable: brute force attacks.\\

The safeness of using ANS based cryptography depends on the way we would use it and the concrete scenario. Having access to both input and output of discussed automata, one usually could deduce the used coding tables - there would be needed some additional protection to prevent that. However, if there was a secure PRNG used to generate them, obtaining the tables does not compromise the key. So initializing PRNG with the cryptographic key and additionally some number (e.g. of data block), would allow to choose many independent encodings for this key.

Beside combining entropy coding with encryption, ANS can be also seen as a cheap and simple building block for stronger cryptographic systems, like a replacement for the memoryless S-box. Example of such cryptography can be: use the PRNG to choose an intermediate symbol probability distribution and two ANS coders for this distribution. Encoding of a bit sequence would be using the first encoder to produce a symbol in this distribution, and immediately apply the second decoder to get a new bit sequence.
\section{Conclusions}
There was presented simpler alternative to arithmetic coding: using a single natural number as the state, instead of two to represent a range. Its multiple ways of use give alternatives for different variants of AC. However, there are some advantages thanks to this simplicity:
\begin{itemize}
  \item instead of complex AC renormalization procedure, we can just transfer bit blocks of known size (once per symbol),
  \item there appears additional variant (tANS): putting coder for large alphabet into a table, combining speed of Huffman coding with compression rate of AC, 
  \item the huge freedom while choosing the exact encoder and chaotic state behavior makes it also perfect to simultaneously encrypt the data or as an inexpensive nonlinear building block of cryptosytems.
\end{itemize}
The disadvantages are:
\begin{itemize}
  \item tANS requires building and storing coding tables: about 1-16kB for 256 size alphabet,
  \item the decoding is in opposite direction to encoding, what requires storing the final state and may be an inconvenience, especially while adaptive applications.
\end{itemize}
While maintaining the state space ($I$), we can freely change between different probability distributions, alphabets or even ABS/ANS variants. For example for complex tasks like video compression, let say the used first 32 coding tables can be for quantized different probabilities for binary alphabet, while succeeding tables would be for static probability distributions on larger alphabets for  various specific contexts, like for quantized DCT coefficients. The final memory cost would be a few kilobytes times the number of tables/contexts.\\

There has remained many research question regarding this novel approach. Especially how to effectively generate good tANS coding tables for given probability distribution (including tuning). We should also add correlated sources to considerations and generally improve theoretical analysis, understanding of behavior. Another research topic regards cryptographic capabilities of this new approach - to understand how to use it, combine with other methods to obtain required level of cryptographic security.

Finally we should understand fundamental questions regarding low state entropy coding automata - is this method optimal, are there maybe more convenient families?

\end{document}